# A Pilot Clinical Study to Investigate the Human Whole Blood Spectrum Characteristics in the Sub-THz Region

Tzu-Fang Tseng[1,2], Borwen You[2], Hao-Cheng Gao[1,2], Tzung-Dau Wang[2,3,4], and Chi-Kuang Sun[1,2,4,5,*]

[1]*Graduate Institute of Photonics and Optoelectronics, National Taiwan University, Taipei, 10617, Taiwan*
[2]*Molecular Imaging Center, National Taiwan University, Taipei, 10617, Taiwan*
[3]*Cardiovascular Center and Division of Cardiology, Department of Internal Medicine, National Taiwan University Hospital, Taipei, 10002, Taiwan*
[4]*College of Medicine, National Taiwan University, Taipei, 10051, Taiwan*
[5]*Research Center of Applied Science and Institute of Physics, Academia Sinica, Taipei, 11529, Taiwan*
[*]sun@ntu.edu.tw

**Abstract:** We have conducted a pilot clinical study to not only investigate the THz spectra of *ex-vivo* fresh human whole blood of 28 patients following 8-hours fasting guideline, but also to find out the critical blood ingredients of which the concentration dominantly affects those THz spectra. A great difference between the THz absorption properties of human blood among different people was observed, while the difference can be up to ~15% of the averaged absorption coefficient of the 28 samples. Our pilot clinical study indicates that triglyceride and red blood cell were two dominant factors to have significant clinically defined negative correlation to the sub-THz absorption coefficients.

## 1. Introduction

Blood is the most important body fluid, containing many kinds of proteins, lactose, blood cells, and salts. Many biochemical variables in blood plasma, such as glucose and total cholesterol, are highly related to diseases such as diabetes and heart diseases. The concentration and molecular outcomes of most biochemical variables can only be examined *ex-vivo*. For example in nowadays, triglyceride in plasma can only be examined by enzymatic colorimetric analysis [1]. In this method, certain enzymes are added into plasma to interact with the examined factor. The formed chemical compound absorbs visible light at a certain frequency, and the concentration of the factor can be then estimated according to the light absorption coefficient of the compound [1]. To avoid the visible light absorption by red blood cells, this method needs blood centrifuge. Recently THz wave has been proved to have strong interactions with many biochemical molecules such as amino acids[2,3], proteins[4,5], and deoxyribonucleic acid [6] (DNA). In aqueous solutions composed of water, THz wave has also been found to be a sensitive tool to investigate the collective motions of water molecules [7-9]. Blood has around 80% water content [10]. As a result, comparing to the existing examination method, THz wave could be a potential candidate to not only investigate the dynamic molecular behaviors in blood, but also the concentrations of some critical biochemical variables in the human whole blood, and even to be applied non-invasively for in vivo wearable device examination [11,12] or in vivo molecular imaging applications [13-16].

    In the past, there were two patents describing non-invasive THz blood examination devices [11,12] and two journal articles investigating the THz dielectric properties of the ex-vivo blood of human and rat [17,18]. However, in the two journal articles only less than 4

samples were studied. Since previous researches have indicated that proteins, ions, and lipids might all modify the THz absorption coefficients in aqueous solutions [4-9], and the concentrations of these factors in blood are with a huge variation by individuals, the investigated THz blood spectra of few samples would be more likely to have high deviations to the spectra of the general population. Conducting a clinical study which is designed to demonstrate the effect of a certain technique on a selected subset of the general population [19] is the most promising way to not only have a representative human blood spectra investigation, but also to find out the dominant factors that strongly affect the THz spectral characteristics of human whole blood.

In this work we have conducted a pilot clinical study to not only investigate the THz spectra of *ex-vivo* fresh human whole blood of 28 patients following 8-hours fasting guideline, but also to find out the critical blood ingredients of which the concentration dominantly affects those THz spectra. A great difference between the THz absorption properties of human blood among different people was observed, while the difference can be up to ~15% of the averaged absorption coefficient. The absorption coefficients in the low sub-THz frequency region of the 28 samples were dominated by the red blood cell count with a significant negative correlation. In the high sub-THz region, the THz absorption coefficients were found to be significantly negative correlated to the concentration of triglyceride. For the other examined factors, no significant correlation was observed, probably due to the dominance of the red blood cell count fluctuation as well as the triglyceride concentration variation. Our study suggests that the two dominant factors of triglyceride concentration and RBC count would need to be well controlled for futuristic correlation investigation to the other biochemical factors in human whole blood.

## 2. Method

*2.1 Experimental Setup*

A portable THz time-domain spectrometer (mini-Z, Zomega Inc.) was placed in the laboratory right next to the cardiac catheterization laboratory of National Taiwan University Hospital, to enable fresh human blood spectrum acquisition immediately after blood extraction from every experimental subject. The portable THz time-domain spectrometer used an external fiber-coupled 1.5 μm pulsed laser with a 140mW average power, with a less than 100fs pulse duration, and with a pulse repetition rate of 100 MHz as the pump source that was split into a pump and a probe beam for THz wave generation and detection. A large-aperture photoconductive antenna was used to generate the broadband THz wave and an electro-optic crystal was used to detect the THz waveform. The system measured the THz spectrum from 0.1 THz to 3 THz. The peak dynamic range was greater than 48dB under the 0.5 second acquisition condition. To improve the signal to noise ratio, in this paper we present the 1-minute waveform by averaging 120 continuous 0.5-second waveforms in sequence numerically. The fluidic-sample chamber for transmission-type measurement was self-designed with 2 polyethylene (PE) windows, one was a flat PE cap and the other was with a 100μm-thick 1.5cm-wide fluidic channel at one side. The PE window and PE cap were sandwiched by two aluminum frames and locked by four nuts, and the injected liquid can thus be sealed. The photos of the chamber with human blood injection are shown in Fig. 1(a), 1(b) and 1(c). All the experiments were operated at room temperature of $24^oC$ with a stable humidity 60%. In each time of the measurement, the reference spectrum of the empty chamber was acquired one minute before the sample injection. Our measurement repeatability was tested by injecting bulk water, with the chamber assembled, fixed, injecting water, and disassembled three times. Figure 2 shows the measured absorption coefficient spectra of water for three measurements. The ratio of the standard deviation to the average value was less than 1% in all frequencies between 0.1-1.2THz. Due to high water absorption, no reliable data was acquired beyond 1.2 THz in this study.

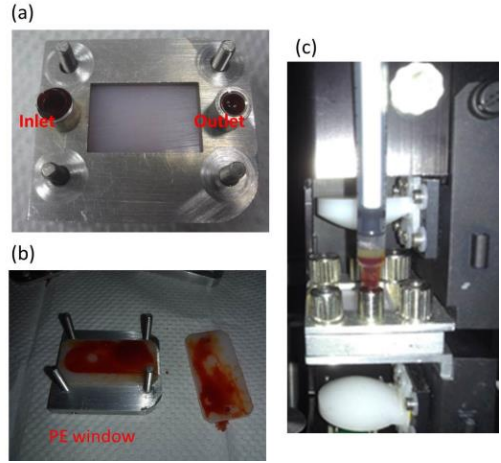

Fig. 1. (a) Self-designed fluidic channel. (b) The PE cap with human blood. (c) The blood injection process, with a portable THz spectrometer.

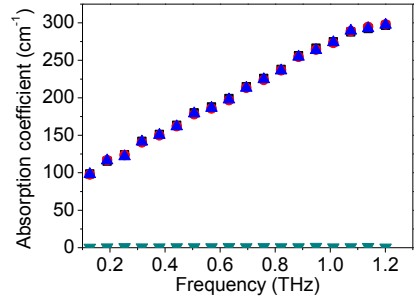

Fig. 2. Three-times measured absorption coefficients of bulk water shown as the top black, red, and blue dots. The bottom dots show their standard deviations.

### 2.2 Clinical Protocol: Blood Spectra Acquisition

This study was conducted according to the Declaration of Helsinki Principles, and the following protocol was approved by the Institutional Review Board of National Taiwan University Hospital. Informed consent was obtained from each subject prior to study entry. The human blood samples were obtained before cardiac catheterization, and all patients followed 8-hour fasting guidelines before the surgery. We started THz spectrum acquisition 3-4 minutes after blood extraction from patients, which was the shortest duration we could reach. 28 extracted blood samples were injected into heparin vacutainer. Heparin is the most common anticoagulant that has no chemical reactions with other contents except anti-thrombin III, and is usually added in blood for examinations [20]. We started to acquire the reference spectra from the empty chamber one minute before blood injection. The spectrum of each blood sample was recorded 2 spectra/second continuously for 1 minute.

### 2.3 Clinical Protocol :Examination items

For the examined variables, different biochemical risk factors were examined within 24 hours after blood extraction including HbA1c (%), Glucose (mg/dl), total protein (g/dl), albumin (ALB) (g/dl), globulin (GLO) (g/dl), total cholesterol (T-CHO) (mg/dl), triglyceride (TG) (mg/dl), urea nitrogen (UN) (mg/dl), creatinine (CRE) (mg/dl), $Ca^{2+}$ (mmol/L), $Na^{2+}$ (mmol/L), and $K^{2+}$ (mmol/L). Different hematological risk factors were examined one day before the cardiac catheterization, including red blood cell (RBC) count (M/μL), white blood cell (WBC) count (K/μL), platelet (PLT) count (K/μL), mean corpuscular hemoglobin (MCH) (pg), and mean corpuscular hemoglobin concentration (MCHC) (g/dL). The correlations between the spectra of all the 28 samples and these factors were analyzed. It is well known that these hematological risk factors do not fluctuate within one day [21]

Since our blood samples were donated from the patients having cardiac catheterization, these samples were more likely to have an abnormal HbA1c level, TG level, or T-CHO level simultaneously. Thus, in this study we have another selected subset with these variables controlled within the normal region: HbA1c <=5.7%, T-CHO level <200mg/dl, and TG level <200 mg/dl [22,23]. With these limitations, 13 samples were selected.

*2.4 Analysis*

To analyze the measured THz dielectric properties of blood, simple calculations were applied and are described below. The spectrum of the THz electric field transmitted through the empty container $E_{ref}(\omega)$ is:

$$E_{ref}(\omega) = E_s(\omega) \cdot T(\omega)_{(PE-air)} \cdot e^{(i\omega \cdot d/c)} \cdot T(\omega)_{(air-PE)} \quad (1)$$

where $E_s(\omega)$ is the electric field of the THz source, $\omega$ is the THz angular frequency, $d$ is the thickness of the fluidic channel, and $c$ is light speed; whereas the electric field transmitted through the container with blood $E_{sample}(\omega)$ is:

$$E_{sample}(\omega) = E_s(\omega) \cdot T(\omega)_{(PE-blood)} \cdot e^{(i\omega \cdot d \cdot n(\omega)/c)} \cdot e^{(-\omega \cdot d \cdot \kappa(\omega)/c)} \cdot T(\omega)_{(blood-PE)} \quad (2)$$

where $n(\omega)$ and $\kappa(\omega)$ are the real and imaginary parts of the refractive index of blood. $T(\omega)_{a-b}$, which is the transmission coefficient from medium a to medium b, has the form of: $\frac{2\overline{n_a}}{\overline{n_a} + \overline{n_b}}$, where $\overline{n_a} = n_a + i\kappa_a$ is the complex refractive index. The THz absorption in air is neglected. The refractive indices of PE and air are 1.5 [24] and 1, respectively, and the absorption of PE is very small and can be neglected [24]. The complex refractive index of blood was first obtained approximately by :

$$\kappa_{blood}(\omega) \approx \left( \frac{\ln\left(|\frac{E_{sample}(\omega)}{E_{ref}(\omega)}|^2\right)}{liquid\ thickness\ (d)} \right) \times c/2\omega, \quad [25] \quad (3)$$

$$n_{blood}(\omega) \approx (phase\ difference\ (\Delta\phi_{sample-ref})) \times c/\omega d + 1$$

Then the exact value was approached by several times of iterations to eliminate the transmission coefficients through interfaces and to keep the blood induced absorption exponential term only. The absorption coefficient $\alpha$ has the form of $2 \cdot \omega \cdot \kappa(\omega)/c$, where the symbols are defined above. The blood absorption coefficients at two frequencies near 270GHz and 820GHz those are far from water vapor absorption lines (558GHz, 753GHz, 989GHz) [26] were chosen for correlation analysis. The bivariate correlation analysis between the examined items and the THz absorption coefficient was done by a Statistical-Product-and-Service-Solutions (SPSS) program [27]. P-value, the probability of the null hypothesis [28], is a factor to judge if the dependence or correlation has a high significance. The THz blood absorption is considered to have a significant correlation with a specific risk factor only if the p-value is less than 0.05 [28]. When p-value was less than 0.05, the correlation factor r [29] is further analyzed. Within the significant correlation range, we further divide the correlation into three category, moderate to strong (|r|>0.5), weak to moderate (0.3< |r| <0.5), and weak (|r|<0.3).

### 3.Result

Figure 3 shows the acquired 28 ex-vivo absorption coefficient spectra of fresh human whole blood with heparin added. The largest absorption coefficient difference between these samples was 17cm$^{-1}$ at 270GHz, which equaled to 15% of the average absorption coefficient (109.4cm$^{-1}$), and was 27cm$^{-1}$ at 820GHz, which equaled to 13% of the average absorption coefficient (201.6cm$^{-1}$). We can observe that the largest absorption coefficient between these samples was comparable to the absorption coefficient difference between water (the top line in Fig. 3) and blood. This strongly indicates that the THz absorption coefficient must be sensitive to some factors in blood.

With our analysis, RBC count is found to have a negative correlation with the absorption coefficient at 270GHz (r=-0.383, two-tail p-value =0.044, sample number=28, shown in Fig. 4(a)), but not at 820GHz (two-tail p-value =0.736, shown in Fig. 4(b)), indicating another possible factor dominating the high sub-THz region over the red blood cell count As a result, we find that for most blood samples having their TG levels within the normal region (<200mg/dl), their absorption coefficients at 820GHz were dominated by TG concentration with a negative correlation (r=-0.437, p=0.037, sample number=23 shown in Fig. 5 (b)), but not at 270GHz (two-tail p-value =0.056, shown in Fig. 5(a)). For other examined factors, no positive or negative correlation with the THz absorption coefficients at both 270GHz and 820GHz was found, such as glucose (two-tail p-value=0.241 at 270GHz and 0.235 at 820GHz), T-CHO (two-tail p-value=0.569 at 270GHz and 0.732 at 820GHz), ALB (two-tail p-value=0.532 at 270GHz and 0.502 at 820GHz), GLO (two-tail p-value=0.4 at 270GHz and 0.555 at 820GHz), Na$^+$ (two-tail p-value=0.131 at 270GHz and 0.588 at 820GHz), PLT (two-tail p-value=0.567 at 270GHz and 0.422 at 820GHz), and HbA1c (two-tail p-value=0.44 at 270GHz and 0.944 at 820GHz). Then, by statistical analysis to the other subset composed of the picked 13 normal samples with HbA1c <=5.7%, T-CHO level <200mg/dl, and TG level <200 mg/dl, we have found that their absorption coefficients also had a negative correlation to the TG concentration at both 270GHz (r=-0.575, two-tail p-value=0.04, shown in Fig. 6(a)) and 820GHz (r=-0.611, two-tail p-value=0.027, shown in Fig. 6(b)), and had no significant correlation to other factors such as glucose (two-tail p-value=0.635 at 270GHz and 0.781 at 820GHz), T-CHO (two-tail p-value=0.618 at 270GHz and 0.243 at 820GHz), ALB (two-tail p-value=0.154 at 270GHz and 0.187 at 820GHz), GLO (two-tail p-value=0.392 at 270GHz and 0.291 at 820GHz), Na$^+$ (two-tail p-value=0.106 at 270GHz and 0.736 at 820GHz), PLT (two-tail p-value=0.187 at 270GHz and 0.054 at 820GHz) and HbA1c (two-tail p-value=0.29 at 270GHz and 0.417 at 820GHz).

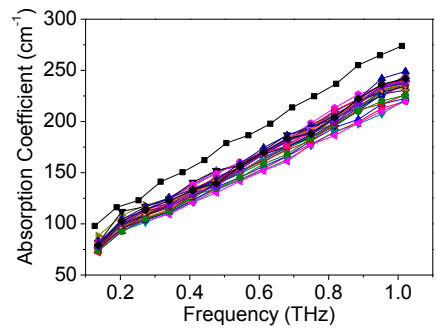

Fig.3. The top line shows the absorption coefficient of water, and the other lines below show the different absorption coefficients of the blood samples measured in the first minute, of the 28 patients with heparin added.

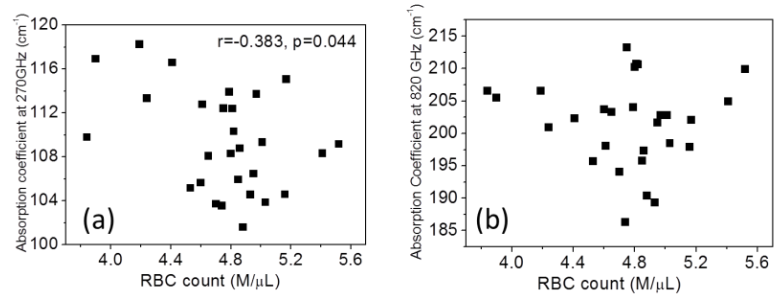

Fig. 4 The THz absorption coefficients of the 28 blood samples versus RBC count. (a) at 270GHz (negative correlation r=-0.383, two-tail p-value =0.044) (b) at 820GHz (two-tail p-value =0.736).

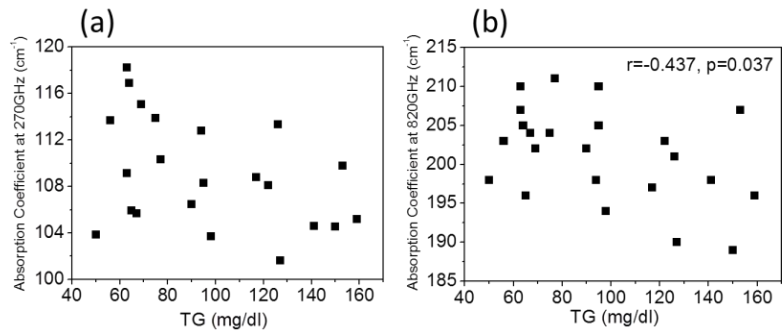

Fig. 5 The THz absorption coefficients of the 23 blood samples with their TG levels less than 200mg/dl (a) at 270GHz (p=0.056) (b) at 820GHz (negative correlation r=-0.437, p=0.037).

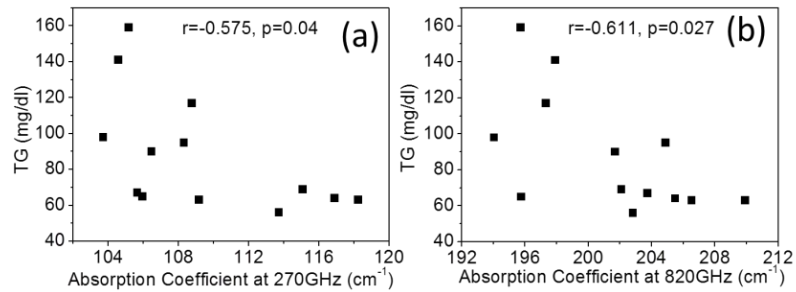

Fig.6 TG concentration versus the absorption coefficients of the blood samples having HbA1c <=5.7%, T-CHO level <200mg/dl, and TG level <200mg/dl (a) at 270GHz (negative correlation r=-0.575, two-tail p-value=0.04) (b) at 820GHz (negative correlation r=-0.611, two-tail p-value=0.027).

This negative correlation to the RBC count at 270GHz corresponds to the fact that RBC occupies the largest volume fraction of blood and varies a lot by individual between 37% to 52% [21], and to previous reports that RBC absorbs less THz power than plasma [17,18]. However, the p-value (0.044) was not very small, which was different from the perfect inverse relation between the RBC volume and the absorption coefficient of the rat blood exhibited in reference [17]. The TG-dominant negative correlation in the THz region has not been described in any previous researches. The most dominant lipids in blood include cholesterol and fatty acids [30]. As a critical blood lipid, a triglyceride is an ester derived from glycerol and three fatty acids [30]. Even though it is reasonable to understand the observation on the negative correlation between TG concentration and THz absorption due to that lipids absorb much less THz power than water, it is however harder to understand the comparatively high p-value for the analysis on T-CHO, due to the fact that both T-CHO and TG are with a similar mass concentration per unit volume of blood (most of the blood samples had their T-CHO within 100mg/dl and 200 mg/dl, and TG within 60mg/dl and 200mg/dl). In doctrinal study, it is known that both TG and cholesterol in the whole blood are carried by lipoproteins, including very-low-density lipoprotein (VLDL), low-density lipoprotein (LDL), high-density lipoprotein (HDL), and chylomicron with a density even lower than VLDL [30]. These four kinds of lipoproteins have different densities and different percentages of TG and cholesterol composition, as summarized in Table 1. As can be seen in Table 1, the lipoprotein having higher TG weight percentage has a lower density. While TG and CHO have the similar mass concentration per unit volume of blood, the volume faction of TG in the whole blood is higher than CHO. Combining with the fact that the TG concentration in blood has a higher variation than T-CHO, these two factors combined might be responsible for our observation that TG dominates the correlation over T-CHO.

In summary, our results have shown that human blood is a very complicated fluid which has the THz spectrum characteristics different from the predicted ones of the simple aqueous solutions. To further study the correlations between the THz spectrum of the whole blood and other examined factors such as proteins and ions, our results have indicated that RBC and TG concentration are the two primary parameters that need to be well controlled.

**Table 1. The density and different percentages of TG and cholesterol composition of different kinds of lipoprotein**

| Lipoprotein | Density (g/mL) | Free Cholesterol + Ester of Cholesterol (weight%) | Triglyceride (weight%) |
|---|---|---|---|
| Chylomicrons | <1.006 | 4 | 85 |
| VLDL | 0.95-1.006 | 19 | 50 |
| LDL | 1.006-1.063 | 45 | 10 |
| HDL | 1.063-1.210 | 17 | 4 |

## 4. Conclusion

In conclusion, a pilot clinical study of the THz human blood spectroscopy was conducted. A great difference between the THz absorption properties of human blood among different people was observed. We have found that TG concentration is the dominant factor to affect the whole blood absorption coefficient in the high sub-THz region and is with a significant negative correlation to the THz absorption coefficient. The RBC count was on the other hand found to dominate the low sub-THz region and is with a significant negative correlation to the THz attenuation constant. For the concentration of the ions including $Na^+$, $Ca^{2+}$, and $K^+$, the glucose level, and the concentration of the total proteins including albumin and globulin, no correlation to the absorption coefficient at the low and high sub-THz frequencies was observed in this clinical study. Clinically, RBC count and TG concentration would need to be well controlled for futuristic THz correlation investigation to the other factors in human blood.

**Acknowledgment**

This project is sponsored by the Ministry of Science and Technology MOST 103-2112-M-002-016-MY3. We also thank engineer Ching-Chang Liao for machining help.